\documentclass[prl,aps,twocolumn,floatfix,showpacs]{revtex4}
\usepackage{graphics}
\begin{document}
\title{Vortices, antivortices and superfluid shells separating Mott-insulating regions} 
\author{Kaushik Mitra, C. J. Williams, C. A. R. S{\'a} de Melo}
\affiliation{Joint Quantum Institute \\ University of Maryland, College Park, MD 20742 \\ 
NIST, Gaithersburg, MD 20899}
\date{\today}

\begin{abstract}
Atomic or molecular bosons in harmonically
confined optical lattices exhibit a wedding
cake structure consisting of insulating (Mott) shells. 
It is shown that superfluid regions emerge between Mott shells
as a result of fluctuations due to finite hopping.
It is found that the order parameter equation in the
superfluid regions is not of the Gross-Pitaeviskii type except near the insulator 
to superfluid boundaries. The excitation spectra
in the Mott and superfluid regions are obtained, and it is
shown that the superfluid shells posses low energy sound modes with spatially dependent sound
velocity described by a local index of refraction directly related to the local superfluid density.
 Lastly, the Berezinskii-Kosterlitz-Thouless 
transition and vortex-antivortex pairs are discussed
in thin (wide) superfluid shells (rings) limited by three (two) dimensional Mott regions. 
\end{abstract}
\maketitle

{\it Introduction:}
The recent experimental discovery of Bose-Mott insulating phases 
in optical lattices has generated
an explosion of research in the ultra-cold atom community (see~\cite{bloch-review-2005} for
a recent review), and has helped to merge two major branches of physics: atomic-molecular-optical and condensed
matter physics. Most experiments thus far have relied on time of flight measurements to infer
the existence of a superfluid to insulator transition~\cite{greiner-2002, stoferle-2004}. However,
very recently, two experimental groups~\cite{MIT-2006, Mainz-2006} have used spatially selective
microwave spectroscopy to probe {\it in situ} the superfluid-to-insulator transition of $^{87}$Rb 
in a three dimensional (3D) optical lattice with a harmonic envelope.
In these experiments, the shell structure of the Bose-Mott insulating states was revealed for very deep lattices. 
Regions of filling fraction $n = 1$ through $n = 5$ ($n = 1$ through $n = 3$) 
were mapped in the MIT~\cite{MIT-2006} (Mainz~\cite{Mainz-2006}) experiment. Their observation
in three-dimensional optical lattices lead to the confirmation of 
the Mott-insulating shell structure consisting of ``Mott plateaus'' with
abrupt transitions between any two sucessive shells as proposed in two dimensional (2D)~\cite{jaksch-1998} 
optical lattices.

\begin{figure} [htb]
\centerline{ \scalebox{0.33}{\includegraphics{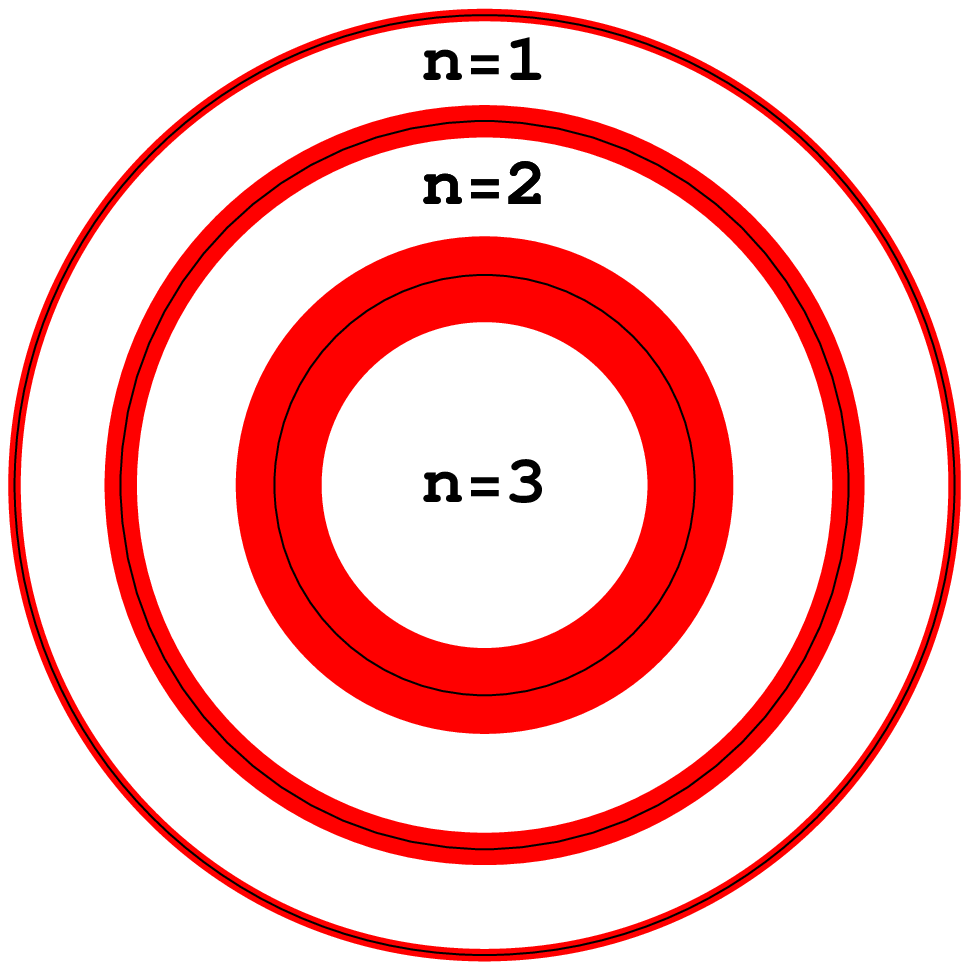} } 
\scalebox{0.50} {\includegraphics{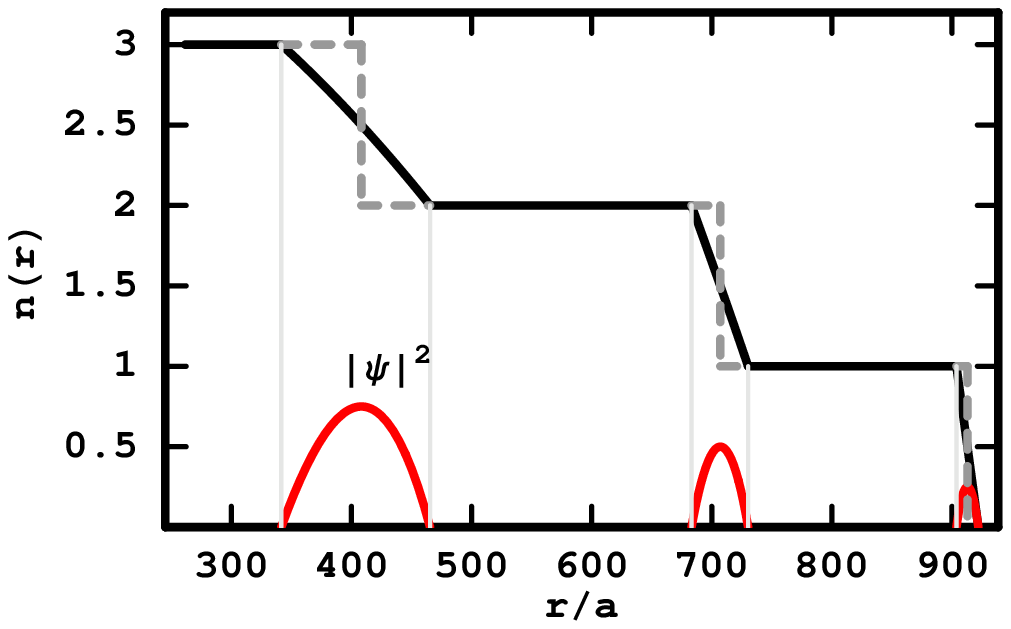} }
}
\caption{ \label{fig:1}
(color online) 
a) Shell structure of Mott and superfluids regions 
in a 2D square optical lattice with harmonic envelope 
as a function of radius $r/a$ for $t \ne 0$.
The superfluid regions are shown in red (gray)
whereas the Mott regions are shown in white. 
The black circles indicate the Mott boundaries $R_{c,n}$ at $t = 0$.
b) The local filling factor $n (\mathbf{r})$ is shown in solid (dashed) lines for $t\ne 0$ 
$(t = 0)$. The red curve (solid gray) shows the local superfluid order parameter $|\psi ({\mathbf r})|^{2}$. 
The parameters are $\Omega= 6 \times 10^{-6} U$,
$ t = 1.25\times 10^{-2} U$ and $\mu = 2.5U$.
}
\end{figure}

One of the next frontier for ultra-cold bosons in optical lattices is the search for
superfluid regions separating Mott-insulating shells. These additional
superfluid regions can be determined through the use of microwave spectroscopy much in the
same way as the Mott shell structure was determined recently~\cite{MIT-2006, Mainz-2006}. 
Thus, in anticipation of the next experimental breakthrough,
we study 2D and 3D optical lattices of atomic or molecular bosons in harmonically confining
potentials, and show that between the Mott regions of filling fraction $n$ and $n+1$,
superfluid shells emerge as a result of fluctuations due to finite hopping.
This finite hopping breaks 
the local energy degeneracy of neighboring Mott-shells, determines
the size of the superfluid regions as shown in Fig~\ref{fig:1}, 
and is responsible for the low energy (sound) and vortex excitations.
In addition, we find that the order parameter equation is not
in general of the Gross-Pitaeviskii type. 
Furthermore, in 3D optical lattices, when superfluid regions are thin (nearly 2D) spherical (or ellipsoidal) shells,
we obtain bound vortex-antivortex excitations below the Berezinski-Kosterlitz-Thouless (BKT) transition
temperature~\cite{berezinski, kosterlitz-thouless} which is different for each superfluid shell.

{\it Hamiltonian:}
To describe the physics of alternating insulating and superfluid shells for atomic
or molecular bosons we use the lattice Bose-Hubbard Hamiltonian with 
a harmonic potential described by
\begin{equation}
\label{eqn:bose-hubbard}
H = -t\sum_{\mathbf{r},\mathbf{a}}c_{\mathbf{r}}^{\dagger}c_{\mathbf{r+a}}+
\frac{U}{2} \sum_{\mathbf{r}}
c_{\mathbf{r}}^{\dagger} c_{\mathbf{r}}^{\dagger} c_{\mathbf{r}} c_{\mathbf{r}} 
-\sum_{\mathbf{r}} \mu_{\mathbf{r}}  c_{\mathbf{r}}^{\dagger}c_{\mathbf{r}} ,
\end{equation}
where $\mu_{\mathbf{r}}= \mu - V(r)$ is the local chemical potential,
$V(r) = \Omega_{\rho} (\rho/a)^2/2 + \Omega_z (z/a)^2/2$ is the harmonically confining potential,
$a$ is the lattice spacing, and $c_{\mathbf{r}}^{\dagger}$ is the creation operator 
for an atom at site $\mathbf{r}$.
Here, $t$ is the hopping parameter and $U$ is the interaction strength.

We introduce a local superfluid order
parameter $\psi_{\mathbf{r}}=\langle c_{\mathbf{r}} \rangle$ that 
produces an effective local Hamiltonian 
\begin{equation}
\label{eqn:bose-hubbard-local}
H_{\mathbf{r}} ^{\textrm{eff}} = 
H_{0,n} (\mathbf{r}) 
 - t \sum_{\mathbf{a}}(c_{\mathbf{r}}\psi^{*}_{\mathbf{r+a}}\
+ c_{\mathbf{r}}^{\dagger}\psi_{\mathbf{r+a}}
-\psi^{*}_{\mathbf{r}}\psi_{\mathbf{r+a}}),
\end{equation}
which is diagonal in the site index 
$\mathbf{r}$ with 
$H_{0,n} (\mathbf{r}) = \frac{U}{2}\hat{n}_{\mathbf{r}}(\hat{n}_{\mathbf{r}}-1)
-\mu_{\mathbf{r}} \hat{n}_{\mathbf{r}}$, where
$\hat{n}_{\mathbf{r}}=c_{\mathbf{r}}^{\dagger}c_{\mathbf{r}}$
is the number operator. For $t = 0$, the Mott shell structure is revealed by 
fixing $ \hat{n}_{\mathbf r} = n$, to obtain the local energy
$E_{0,n} (\mathbf{r}) = U n(n-1)/2 - \mu_{\mathbf{r}} n$, when $(n - 1)U < \mu_{\mathbf{r}} < nU$. 
Since $E_{0,n+1} (\mathbf{r}) -  E_{0,n} (\mathbf{r}) = nU -  \mu_{\mathbf{r}}$,
the change from a Mott shell with filling fraction $n$ to $n + 1$ occurs
at the degeneracy condition $\mu_{\mathbf{r}} = nU$, which for a spherically symmetric potential
happens at $R_{c,n} = a \sqrt{ 2 (\mu - nU)/\Omega}$. 
The relation $\mu_{\mathbf{r}} = nU$ determines the shape and size
of the boundary between the $n$ and $n+1$ shells.
However, as shown next, near this region of degeneracy 
fluctuations due to hopping introduce superfluid shells.

{\it Nearly Degenerate Perturbation Theory:}
We focus our attention now on the Mott regions with integer boson
filling $n$ and $n+1$ and the superfluid shell that emerges 
between them. In the limit where $U\gg t$ we can restrict our Hilbert
space to number-basis states $|n\rangle$ and $|n+1\rangle$ at each
site. Any contribution of other states to the local energy will be of the order of $t^{2}/U$.
The hopping term in Eq.~(\ref{eqn:bose-hubbard-local})
affects the ground state energy of the system by removing
the local degeneracy of $E_{0,n+1} (\mathbf{r})$ 
and $E_{0,n} (\mathbf{r})$ at $\mu_{\mathbf{r}} = n U$.
To illustrate this point, we make a continuum
approximation and perform a Taylor expansion of $\Psi ( \mathbf{r} + \mathbf{a} )$
to second order in $a$, provided that $r \gg a$. This approximation is
not really necessary, but allows substantial analytical insight leading 
to 

\begin{equation}
\label{eqn:h-matrix}
H_{\mathbf{r}}^{\textrm{eff}}=
\left(\begin{array}{cc}
E_{0,n} (\mathbf{r}) + \Lambda(\mathbf{r}) & -\sqrt{n+1}\Delta (\mathbf{r})\\
-\sqrt{n+1}\Delta^{*} (\mathbf{r}) & E_{0, n+1}(\mathbf{r}) + \Lambda(\mathbf{r})
\end{array}\right),
\end{equation}
where $\Lambda(\mathbf{r})=\frac{1}{2}(\Delta(\mathbf{r})\psi^{*}(\mathbf{r})+cc)$
and $\Delta(\mathbf{r})= t (z\psi(\mathbf{r})+a^{2}\nabla^{2}\psi(\mathbf{r}))$.
Here, $z$ is the coordination number which depends on 
dimension $d$. Notice that $ t \ne 0$ has two effects. First, it changes the
local energies $E_{0, n}(\mathbf{r})$ and $E_{0, n+1}(\mathbf{r})$ of the Mott shells $n$ and $n+1$
through $\Lambda(\mathbf{r})$. Second, it mixes the two Mott regions through the off-diagonal
term $\sqrt{n+1}\Delta (\mathbf{r})$ and its hermitian conjugate. Thus, the physics near the boundary
between the $n$ and $n+1$ Mott regions is described
by an effective local two-level system with diagonal ($\Lambda(\mathbf{r})$) and
off-diagonal ($\sqrt{n+1}\Delta (\mathbf{r})$) perturbations.

The eigenvalues of Eq.~(\ref{eqn:h-matrix}) are given by, 
$$
\label{eqn:eigenvalues}
E_{\pm} (\mathbf{r}) 
= E_s (\mathbf{r}) \pm\sqrt{ ( E_d (\mathbf{r}))^2
+ (n+1)\left|\Delta(\mathbf{r})\right|^{2}},
$$

where $E_s(\mathbf{r}) = \left[ E_{0, n+1}(\mathbf{r})+ E_{0, n}(\mathbf{r})\right]/2 + \Lambda(\mathbf{r})$
is proportional to the sum of the diagonal terms, and
$E_d(\mathbf{r}) = 
\left[ E_{0, n+1}(\mathbf{r})-  E_{0, n}(\mathbf{r})\right]/2 = ( nU - \mu_{\mathbf{r}} )/2$
is proportional to their difference. 
Notice that $E_{-} (\mathbf{r})$ is the lowest local energy, leading to 
the ground state energy 
$ \mathbf{E} = \frac{1}{L^d}\int d \mathbf{r}E_{-} (\mathbf{r}).$

{\it Order Parameter and Compressibility:}
The order parameter equation (OPE) is determined 
by minimization $E$ with respect to $\psi^{*}(\mathbf{r})$
leading to 
\begin{equation}
\label{eqn:order-parameter}
\Delta(\mathbf{r})
-\frac{(n+1) t (z+a^{2}\nabla^{2})\Delta(\mathbf{r})}
{2\sqrt{\left(\frac{E_d(\mathbf{r})}{2}\right)^{2}+
(n+1)\left|\Delta(\mathbf{r})\right|^{2}}}=0.
\end{equation}
%
%
%
Notice that the OPE is not of the Gross-Pitaeviskii (GP) type,
since the superfluid regions emerge from local fluctuations between 
neighboring Mott shells.
The zeroth order solution of this equation leads to the  
spatially dependent order parameter 
\begin{equation}
\left|\psi(\mathbf{r})\right|^{2} = \frac{n+1}{4}
-\frac{\left(nU - \mu_{\mathbf{r}}\right)^{2}}{4z^{2} t^{2}(n+1)}\label{psi}.
\end{equation}
Since $\left|\psi(\mathbf{r})\right|^{2} \ge 0$, hence $| n U - \mu_{\mathbf{r}}| \le (n+1) z t$, and 
the saddle point inner $R_{n,-}$ and outer $R_{n, +}$ radii for the superfluid shell 
between the $n$ and $n + 1$ Mott regions is obtained by setting 
$\left|\psi(\mathbf{r})\right|^{2} = 0$ 
leading to 
\begin{equation}
\label{eqn:radii}
R_{n, \pm}= R_{c,n} \sqrt{ 1 \pm
\frac{2 z t (n+1)}{\Omega} \frac{a^2}{R_{c,n}^2} }.
\end{equation}

for a spherically symmetric harmonic potential $V(r) = \Omega (r/a)^2/2$, where
$R_{c,n}$ is defined above. This relation
shows explicitly that $t$ splits the spatial degeneracy of the $n$ and $n+1$ Mott shells at
$r = R_{c,n}$ or $(\mu_{\mathbf r} = nU)$
by introducing a superfluid region of thickness $\Delta R_n = R_{n,+} - R_{n,-}$. 
In the case of  non-spherical harmonic potential $V(r) = \Omega_{\rho} (\rho/a)^2/2 + \Omega_z (z/a)^2/2$
the shell regions are ellipsoidal instead of spherical.
Notice that $\Delta R_n$ depends
strongly on filling fraction $n$, chemical potential $\mu$ through $R_{c,n}$ and
the ratio $zt/\Omega$.

In addition, the local filling fraction 
\begin{equation}
n(\mathbf{r}) = -\frac{\partial E_-(\mathbf{r})}{\partial
\mu}=n+\frac{1}{2}-\frac{n U - \mu_{\mathbf{r}}} {2z t (n+1)}
\end{equation}
in the same region interpolates between $n+1$ for $r \lesssim  R_{n, -}$ and
$n$ for $r \gtrsim  R_{n, +}$, while the chemical potential $\mu$ is fixed by the total number
of particles $N={\int d \mathbf{r} n(\mathbf{r})}$. 
In Fig.~\ref{fig:1}, $n(\mathbf{r})$, $\left|\psi(\mathbf{r})\right|^{2}$, and
$R_{n, \pm}$ are shown for the Mott and superfluid regions.
The local compressibility 
\begin{equation}
\label{eqn:compressibility}
\kappa (\mathbf {r})=\frac{\partial n(\mathbf{r})}
{\partial \mu }=\frac{1}{2 z t (n+1)}
\end{equation}
of the superfluid shells is non-zero, 
in contrast to the incompressible ($\kappa = 0$) $n$ and $n + 1$ Mott shells
for  $r < R_{n, -}$ and $r > R_{n, +}$, respectively. The local compressibility 
indicates the presence of large (small) excitation energies
in the Mott (superfluid) regions, as discussed later. 

{\it Gross-Pitaeviskii behavior at the edges:} 
We note in passing that only near the edges of the superfluid region
(where $r\approx R_{n, \pm}$ and  $\psi(\mathbf{r})\approx 0$) 
a direct expansion of the OPE (Eq.~\ref{eqn:order-parameter}) 
leads to the effective Gross-Pitaeviskii equation
\begin{equation}
\label{eqn:Gross-Pitaeviskii}
\left( -\frac{\hbar^{2}}{2m_{\rm eff}}\nabla^{2}
+ V_{\rm eff} (\mathbf{r}) + 
g_{\rm eff} \left|\psi(\mathbf{r})\right|^{2} \right) \psi(\mathbf{r})= 0.
\end{equation}
Here, $m_{\rm eff} = 1/2a^{2} t$ is exactly the boson band mass, 
$V_{\rm eff} (\mathbf{r}) =  |n U - \mu_{\mathbf{r}}|/(n + 1) - zt$ is 
the effective potential, and
$g_{\rm eff} = 2zt/(n + 1)$ is the effective interaction.
Notice that $V_{\rm eff} (\mathbf{r}) \le 0$ and vanishes at the boundaries
$R_{n,\pm}$ of the superfluid region since $|n U - \mu_{\mathbf{r}}| =  zt(n+1)$ there. 
Furthermore, notice $g_{\rm eff} = zt/(n + 1)$ is small in comparison to $U$, indicating
that the superfluid near the edges is weakly interacting, and more so as the Mott index $n$ increases.
When $t \to 0$ $(m_{\rm eff} \to \infty)$ then $\left|\psi(\mathbf{r})\right|^{2} = -V_{\rm eff}/g_{\rm eff}$
leading to the correct limiting behavior of Eq.~\ref{eqn:order-parameter} near $r\approx R_{n, \pm}$.
Next, we discuss the excitation spectrum in the Mott and superfluid regions.

{\it Excitation Spectrum:} The excitation spectrum
in the Mott and superfluid shells can be obtained using 
the functional integration method~\cite{stoof-2001}
for the action ($\hbar = k_B = 1$, $\beta = 1/T$)
\begin{equation}
S[c^\dagger,c] = 
\int_{0}^{\beta}d\tau\sum_{\mathbf{r}}
\left[ 
c_{\mathbf{r},\tau}^{\dagger} {\partial_\tau} c_{\mathbf{r}, \tau}\nonumber  + H 
\right]
\end{equation}
leading to the partition function 
$Z = \int {\cal D} c^\dagger {\cal D} c \exp{- S[c^{\dagger},c]}$.  
In each Mott shell (away from the superfluid region) we introduce a Hubbard-Stratanovich field
$\Psi$ to take into account fluctuations due to the presence of finite hopping, and integrate out
the bosons $(c^\dagger,c)$ leading to an effective action 
$$
S_{\rm eff}[\Psi^{\dagger},\Psi] = 
\int d\mathbf{r}\sum_{i\omega,\mathbf{kk'}}
\Psi_{i\omega,\mathbf {k}} \Psi_{i\omega,\mathbf {k'}}^{*}  e^{ {i(\mathbf{k}-\mathbf{k'})\cdot\mathbf{r}} }
G_\mathbf{kk'}^{-1} (i\omega, \mathbf r)
$$
to quadratic order in $\Psi^{\dagger}$ and $\Psi$, where
$\mathbf{k}$, $\mathbf{k'}$ are momentum labels, $\omega$ are Matsubara frequencies,
and 
$$
G_\mathbf{kk'}^{-1} (i\omega, \mathbf r) = 
\epsilon_{\mathbf{k'}} 
\left[ 1 + 
\epsilon_\mathbf{k} \left( \frac{n+1}{i\omega - E_1 (\mathbf{r})}  - \frac{n}{i\omega - E_2 (\mathbf{r})} \right) \right], 
$$ 
with $E_1 (\mathbf{r}) = nU - \mu_{\mathbf{r}}$, $E_2 (\mathbf{r}) = (n-1)U -\mu_{\mathbf{r}}$,
and $\epsilon_{\mathbf{k}}=2t\sum_{j=1}^{d}\cos(k_{j}a)$.
The poles of $G_\mathbf{kk'} (i\omega, \mathbf r)$ are found upon the analytical continuation
$i \omega = \omega + i\delta$, leading to the local excitation energies
$$
\omega_{\pm}=
-\mu_{\mathbf{r}}+\frac{U}{2}(2n-1)-\frac{\epsilon_{\mathbf{k}}}{2}
\pm\frac{1}{2}\sqrt{\epsilon_{\mathbf{k}}^{2}-(4n+2)\epsilon_{\mathbf{k}}U + U^{2}}
$$
%
where the $+$$(-)$ sign labels quasiparticle (quasihole) excitations.
The energy to add a quasiparticle is $ E_{qp} = \omega_+$, while 
the energy to add a quasihole is $E_{qh} = -\omega_-$.

\begin{figure} [htb]
\centerline{ \scalebox{0.40} {\includegraphics{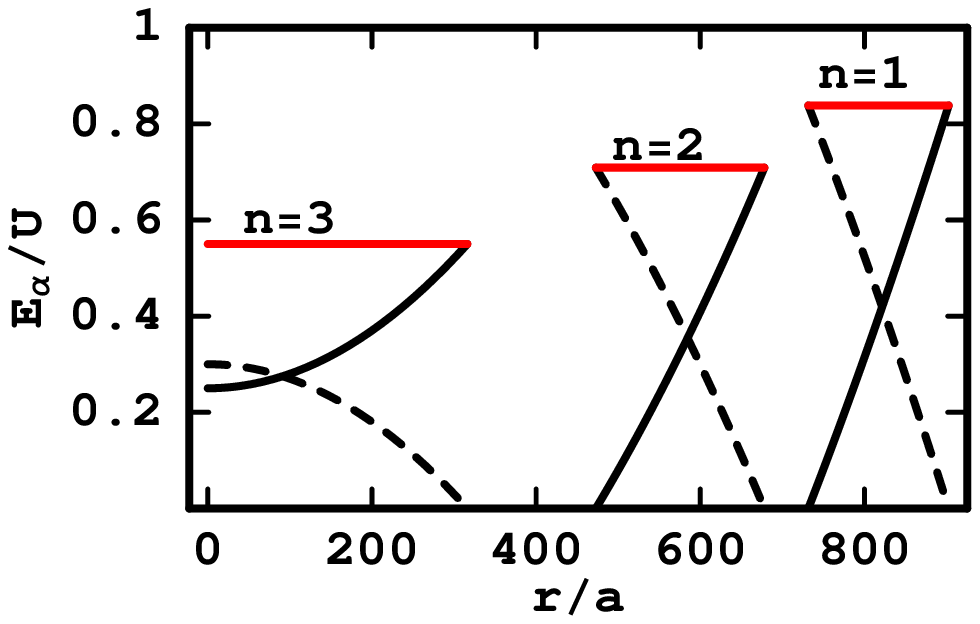}} 
\scalebox{0.40} {\includegraphics{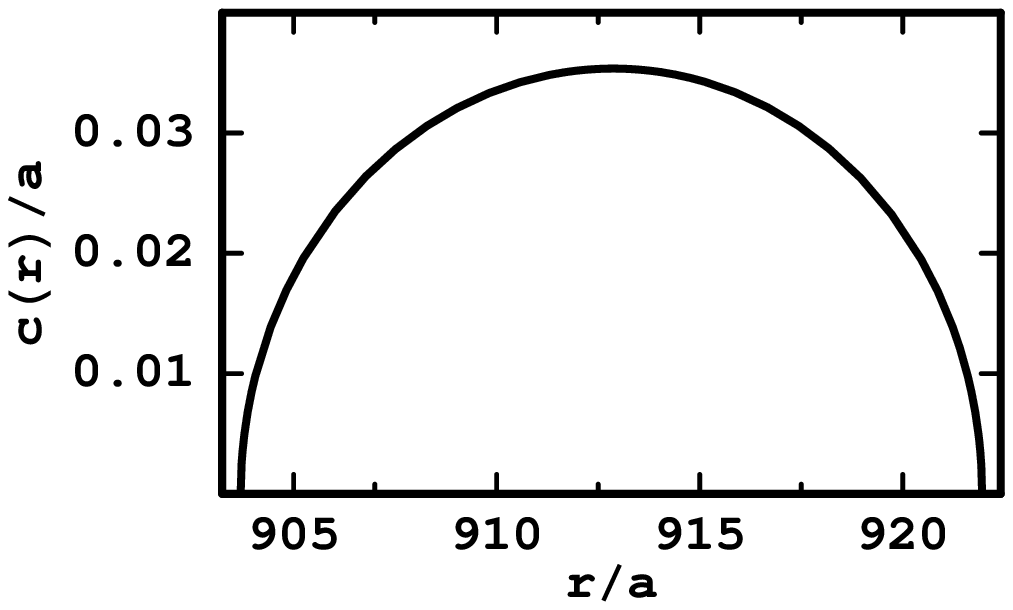}} } 
\caption{\label{fig:2} (color online)
a) Quasiparticle $E_{qp}$ (solid line), quasihole $E_{qh}$ (dashed line),and Mott gap $E_g$ (red) 
energies for $\mathbf{k}= 0$ versus $r/a$  
b) sound velocity for the outermost superfluid ring versus $r/a$. 
Same parameters as in Fig.~\ref{fig:1}.
}
\end{figure}

Figure~(\ref{fig:2})a shows the quasiparticle and quasihole energies
of the Mott phase as a function of position $\mathbf{r}$. The
energy cost to create a quasiparticle (quasihole) is minimum (maximum) at the trap
center and increases (decreases) radially, while the Mott-Hubbard gap $E_{g} = E_{qp} + E_{qh}$
is large and independent of $\mathbf{r}$. 

The excitation spectrum in the
superfluid region can be calculated by a similar method. 
First, we introduce the Hubbard-Stratanovich field $\psi$ which now corresponds to the
order parameter in the superfluid region, second we use an amplitude-phase representation
$\psi (\mathbf{r}, \tau) =  |\psi (\mathbf{r}, \tau)| \exp{[i\varphi(\mathbf{r}, \tau)]}$
and we apply the nearly degenerate perturbation theory described earlier to integrate out
the boson fields $c^\dagger$ and $c$. Thus, we obtain 
the phase-only effective action 
\begin{equation}
\label{eqn:action-superfluid}
S_{\rm eff} = \frac{1}{2 L^d} \int d\mathbf{r}d\tau 
\left[ 
\kappa (\partial_{\tau} \varphi)^2 + \rho_{ij} \partial_i \varphi \partial_j \varphi
\right]
\end{equation}
to quadratic order in the phase variable for the superfluid region between the $n$ and $n+1$ Mott shells. 
Here, we assumed that $|\psi (\mathbf{r}, \tau)|$ is $\tau$-independent
at the saddle point. The coefficient $\kappa$ is the compressibility of the superfluid 
described in Eq.~\ref{eqn:compressibility},
and 
\begin{equation}
\label{eqn:superfluid-density}
\rho_{ij} = 2 t a^2 |\psi (\mathbf{r})|^2 \delta_{ij} = \rho_s (\mathbf{r}) \delta_{ij}
\end{equation}
is the local superfluid density tensor~\cite{footnote}, which is diagonal and vanishes at the Mott boundaries $R_{n,\pm}$. 
The resulting wave equation has the form
\begin{equation}
\label{eqn:wave-equation}
\partial_\tau^2 \varphi - \partial_i \left[ \frac{\rho_s (\mathbf{r})} {\kappa} \partial_i \varphi \right] = 0, 
\end{equation}
leading to a local sound velocity  
$c ({\mathbf r}) = \sqrt{\rho_s (\mathbf{r})/ \kappa}$, 
which in terms of the order parameter reads
$c ({\mathbf r}) = 2 \sqrt{(n+1)z} t a |\psi (\mathbf{r})|$. The local speed of sound has it
maximal value at $|\psi (\mathbf{r})|_{max} = (\sqrt{n+1})/2$, and the superfluid region behaves
as a medium of continuous index of refraction $\chi  ({\mathbf r}) = c_{max}/c ({\mathbf r}) 
= \sqrt{n+1}/(2 |\psi (\mathbf{r})|)$. Notice that $\chi  ({\mathbf r}) \to \infty$ at 
the Mott boundaries where $|\psi (\mathbf{r})| = 0$, indicating that the sound waves of the
superfluid do not propagate into the Mott regions. A plot of the local sound velocity is shown 
in Fig.~\ref{fig:2}b for the superfluid region between the $n = 1$ and $n =0$ Mott shells.
From the phase-only effective action for the superfluid region, we can also investigate
the vortex excitations, which are discussed next.

{\it Vortices:} Next, we explore vortex solutions in two cases where spontaneous vortex-antivortex pairs
can appear as indicators of the 
Berezinski-Kosterlitz-Thouless (BKT) transition~\cite{berezinski, kosterlitz-thouless}.
Case I corresponds to a 3D system, where the superfluid regions are very thin 
$\Delta R_n \ll R_{c,n}$, leading to essentially a two-dimensional superfluid in curved space. 
Case II corresponds to a 2D system, where the superfluid regions are 
thick rings $\Delta R_n \sim R_{c,n}$, leading to essentially a two-dimensional superfluid subject
to boundary conditions. 

In a flat space two-dimensional system stationary vortex solutions 
must satisfy $\oint\nabla\varphi\cdot d\mathbf{l}=2\pi m$,
where $m = \pm1,\pm2,...$ is the vorticity (topological charge) and
$\nabla\varphi$ is the superfluid velocity. The standard vortex solution
in cylindrical coordinates is $\nabla{\varphi} = m \hat{\theta}/r$,
and the corresponding free energy per unit volume is
${\cal F}= \frac{1}{2 L^d}\int d\mathbf{r}\rho_s(\mathbf{r})(\nabla\varphi)^{2}$.
This situation is analogous to a two-dimensional linear
dielectric material where the displacement field
is $\mathbf{D}=\nabla\varphi\times\hat{z}=\epsilon(\mathbf{r})\mathbf{E}$,
with dielectric function $\epsilon(\mathbf{r})=1/\left[ 2\pi\rho(\mathbf{r})\right]$.
Notice that the dieletric function diverges at superfluid boundaries,
since $\rho_s (\mathbf{r}) \to 0$ there.
In this language ${\cal F}$ is identical to the electrostatic
energy per unit volume $U_{\textrm{el}} = \frac{1}{2 L^d}\int d\mathbf{rD}\cdot\mathbf{E}$.
In general the solutions for several vortices (antivortices) can be obtained from 
$\nabla \wedge {\nabla \varphi} = 2\pi {\hat \mathbf{z} }\sum_i m_i \delta ({\mathbf r} - \mathbf{r}_i)$,
where is the location of the vortex (or antivortex) of vorticity $m_{i}$.

\begin{figure} [htb]
\centerline{ \scalebox{0.45} {\includegraphics{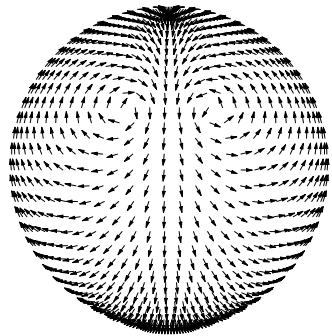}} } 
\caption{\label{fig:3}
Three-dimensional view of a vortex-antivortex pair in 2D superfluid shell separating two 3D Mott regions.
}
\end{figure}

In case I the superfluid state appears below 
$T_{BKT} \approx \pi {\widetilde \rho}_s ( \mathbf{r} = \mathbf{R}_{c,n} )/2$,
where ${\widetilde \rho}_s = \rho_s/a^2$ has dimensions of energy. 
In this limit ${\widetilde \rho}_s ( \mathbf{r} = \mathbf{R}_{c,n} ) = (n+1)t/2$ and
the critical temperature of the superfluid shell between the 
$n$ and $n+1$ Mott regions depends on the index $n$. Notice, however that such estimate
is reasonable only when the shells are sufficiently large. The
solution for a vortex-antivortex (VA) pair in curved two-dimensional space is  
$$
\label{eqn:vortex-antivortex}
\varphi(r = R_{c,n},\theta,\phi)=
\arctan\left(\frac {4 b R_{c,n}\tan\left(\frac{\pi-\theta}{2}\right)\sin(\phi)}
{b^2 - 4 R_{c,n}^2 \tan^2 \left(\frac{\pi-\theta}{2}\right)}\right)
$$
where $R_{c,n} = a \sqrt{ 2 (\mu -nU)/\Omega}$ is the radius of the superfluid shell,
and $b$ is the VA size.
A three-dimensional view of the velocity field $\nabla \varphi (\theta,\phi)$ is shown
in Fig.~\ref{fig:3}. When the superfluid shell has a small
thickness $\Delta R_n$ then $T_{BKT} \approx  \pi t (n+1) \Delta R_n/6 a $, 
while the vortex-antivortex pair has an approximate 
solution of the same form as above which interpolates between $\varphi( r = R_{n,-}, \theta,\phi)$ and 
$\varphi( r = R_{n,+}, \theta,\phi)$.

In case II the superfluid regions are rings bounded by $R_{n,-}$ and $R_{n,+}$, and one can 
use the Coulomb gas analogy described above, conformal mapping techniques and proper boundary
conditions to obtain vortex-antivortex solutions.
The creation of vortex-antivortex 
pairs are energetically quite costly when $\Delta R_n \ll R_{c,n}$, due to strong
confinement effect of the boundaries,
thus we do not expect a BKT-type superfluid transition to occur until 
$\Delta R_n$ is substantially large ($\sim R_{c,n}$). 
Only in this limit, we expect a BKT transition with
$T_{BKT} \approx {\pi \langle {\widetilde \rho}_s \rangle/2}$, 
where $\langle {\widetilde \rho}_s \rangle/2$ is the surface area average of 
${\widetilde \rho}_s ({\mathbf{r}})$.

{\it Conclusions:} 
We studied 2D and 3D optical lattices of atomic or molecular bosons in harmonically confining
potentials, and showed that between the Mott regions of filling fraction $n$ and $n+1$,
superfluid shells emerge as a result of fluctuations due to finite hopping.
We found that the presence of finite hopping breaks 
the local energy degeneracy of neighboring Mott-shells, determines
the size of the superfluid regions as shown in Fig~\ref{fig:1}, 
and is responsible for low energy (sound) and vortex excitations. 
In addition, we obtained an order parameter equation which is not
in general of the Gross-Pitaeviskii type.
Furthermore, we obtained bound vortex-antivortex solutions (as shown in Fig~\ref{fig:3}
below the Berezinski-Kosterlitz-Thouless (BKT) transition
when superfluid regions are thin (nearly 2D) spherical (or ellipsoidal) shells.
The emergence of these superfluid regions should be detectable experimentally using
microwave spectroscopy~\cite{MIT-2006, Mainz-2006}.

We thank NSF (DMR 0304380, and PHY 0426696) for financial support. We also wish to thank F. W. Strauch, Eite Tiesinga, Charles Clark and Ian Spielman for discussions.


\begin{thebibliography}{1}
\bibitem{footnote} The superfluid density tensor has additional corrections beyond the saddle
point proportionality to $|\psi (\mathbf{r})|^2$. These corrections involve gradients 
and laplacians of $|\psi (\mathbf{r})|$.
\end{thebibliography}
\end{document}